\documentclass[9pt, twocolumn]{IEEEtran}
\pdfoutput=1

\usepackage[scale=0.85]{geometry}
\usepackage[draft, margin=false, inline=true, index=true]{fixme}
\fxsetup{theme=color}
\usepackage{multirow}
\usepackage{amsfonts}
\usepackage{subfig}
\usepackage{booktabs}
\usepackage{graphicx}
\usepackage[hyphens]{url}
\usepackage[numbers]{natbib}

\usepackage{amsmath}

\usepackage{url}

\usepackage{color,soul}

\begin{document}

\linespread{1}
\title{Platform-Agnostic Steal-Time Measurement\\in a Guest Operating System}
\author{\IEEEauthorblockN{Javier~Verd\'u, Juan Jos\'e Costa, Beatriz Otero, Eva Rodriguez, Alex~Pajuelo, Ramon Canal}
\IEEEauthorblockA{\\Department of Computer Architecture, Universitat Polit\`{e}cnica de Catalunya\\
Barcelona, Spain\\
Email: \{jverdu, jcosta, botero, evar, mpajuelo, rcanal\}@ac.upc.edu}}

\maketitle

\thispagestyle{empty}

\begin{abstract}
\emph{Steal time}  is a key performance metric for applications executed in a virtualized environment. \emph{Steal time} measures the amount of time the processor is preempted by code outside the virtualized environment.  This, in turn, allows to
compute accurately the execution time of an application inside a virtual machine (i.e. it eliminates the time the virtual machine is suspended).
Unfortunately, this metric is only available in particular scenarios in which the host and the guest OS are tightly coupled. Typical examples are the Xen hypervisor and Linux-based guest OSes. In contrast, in scenarios where the \emph{steal time} is not available inside the virtualized environment, performance measurements are, most often, incorrect.

In this paper, we introduce a novel and platform agnostic approach to calculate this
\emph{steal time} within the virtualized environment and without the
cooperation of the host OS. The theoretical execution time of a \emph{deterministic}
microbenchmark is compared to its execution time
in a virtualized environment. When factoring in the virtual machine load, this solution -as simple as it is-
can compute the \emph{steal time}. The preliminary results show that we are able to
compute the load of the physical processor within the virtual machine with high accuracy.

\end{abstract}



\pagestyle{empty}

\section{Introduction}
\label{sec:intro}
Server consolidation (i.e. running different virtual machines (VMs) in the same
physical machine and sharing its resources) is a cost-effective solution
for datacenters~\cite{crosby2006virtualization}.
These VMs are handled by an underlying hypervisor that implements a time-sharing scheduler to use physical resources. This scheduler assigns a quantum to a given VM. When the quantum expires the hypervisor preempts the VM, suspending its execution, and scheduling a different one. The time between the suspension point and the time the same suspended VM is restarted again is known as~\emph{steal time}. 

Services that run in clouds, like Netflix, a large online video provider, monitor cpu steal metric to detect contention with other collocated virtual machines~\cite{Netflix2014steal}. If so, Netflix kills the virtual machine and recreates it on maybe a different physical machine.

Different cloud providers use different hypervisors, such as Xen in Amazon Web Services (AWS) and Hyper-V in Microsoft Azure, as well as guest VMs based on
different virtualization techniques, mainly with a para-virtualization API (PV)
or without it (also known as hardware assisted virtualization: HVM). Nevertheless, major cloud providers, such as AWS, are moving to mainly use HVM technology~\cite{AWS2018virtualization}.

	In a para-virtualized environment, in which there is a collaboration between the
	guest operating system (OS) and the underlying hypervisor, the
	\emph{steal time} metric is accessible within the VM. Knowing the steal time helps to
	differentiate if a performance problem for a given application comes from the same application or from the surrounding execution environment.

	Without this Hypervisor-OS collaboration there is no such distinction in reported execution time. 
	Consequently, steal time in the guest OS is
	accounted as normal execution time. Therefore
	there is no way to distinguish if performance problems of applications are
	due to the applications themselves or to events from outside the VM.




	Being able to detect this stolen CPU situation inside the VM enables
	smarter VM management such as moving the application to another physical
	machine. Currently this information is only accessible in 
	scenarios where there exist a tight cooperation between the guest OS and the underlying
	hypervisor. For example, in Linux environments running on top of Xen, there is an
	available \emph{steal time} statistic that accounts the percentage of
	time being stolen by code outside the VM. However, if the guest OS is Microsoft Windows\texttrademark based, the statistic cannot be updated, unless the VM runs on top of a Hyper-V hypervisor.

In this paper, we present a novel approach to compute the
\emph{steal time} of a VM within the same VM, by measuring the difference
in the execution time of a deterministic microbenchmark.
We obtain the theoretical execution time of this microbenchmark as a reference
that is later compared with its
execution time inside the VM. The difference between both measurements, in conjunction with the load of the VM,
results in the \emph{steal time}. The main benefit of our approach is that it is
platform agnostic. On one hand, it does not depend on the host OS, or hypervisor, nor on the
guest OS. On the other hand, our proposal works on both PV and HVM environments.
Preliminary results show that this approach is able to detect the load pattern
of the physical CPU within a VM. From this obtained pattern, we derive the \emph{steal time}.

The structure of the paper is as follows. In Section~\ref{sec:background}, we provide a short state of the art. Section~\ref{sec:algorithm}, depicts the main idea of the paper. In section~\ref{sec:methodology}, we explain how to prepare the environment to measure the \emph{steal time} which results are shown in Section~\ref{sec:results}. We also present the future work in Section~\ref{sec:futurework}. Finally, we conclude in Section~\ref{sec:conclusions}.


\section{Related work}
\label{sec:background}
Steal time is a key metric to detect host contention but, in the platforms
where this metric is not available, we need to consider alternative approaches 
like observing differences in the execution
time of the virtualized application\cite{casale2013feasibility}. Our approach is based on this work, but
we remove the requirement to know your application execution time, and instead, we
use a deterministic code to get a general approach.

On one hand, HVM isolates the running virtual machine from other VMs in the
physical machine, and therefore the VM is unaware of the steal time. As far as we know,
there is no approach able to obtain the \emph{steal time} in this scenario, being ours the first one
dealing with this problem.

PV, on the other hand, allows a cooperation between the guest OS and the hypervisor,
making some guest OS statistics accessible like \emph{steal time}. Xen\cite{Barham2003Xen} and  KVM \cite{kivity2007kvm} are the typical examples of this scenario in which the VM knows when it is suspended, and recomputes the statistics accordingly. Few studies have been done about the impact of steal time on benchmark preformance. Schad\cite{Schad2015Understanding} measures steal time in different types of AWS servers by using a Linux performance measurement command, but unfortunately the data is not available under Windows operating system.

The problem with these approaches is that they assign steal time at virtual CPU granularity, without knowing
the percentage of time lost by each virtual task. This problem can be solved
by gathering steal time inside a VM and
distributing it to the running threads\cite{Hofer2015sampling}. This method requires that the underlying
hypervisor offers a steal time statistic inside the VM, which is sampled at fixed intervals to
calculate how the inner running threads are affected. Our approach enhances
this work since the VM is agnostic of the underlying hypervisor and does not require that the \emph{steal time} statistic is visible within the VM.

Another method to link steal time to specific running processes in the guest OS requires
sampling the hypervisor behavior (VM~enter and VM~exit) in the host,
characterizing the \emph{steal} state,  and to
correlate this state with the guest processes to obtain the desired steal
time\cite{yamamoto2016execution}. Since this approach queries the hypervisor for given events, it can only be applied to particular virtualization scenarios.

Chen et al.~\cite{Chen2010XenHVMAcct} modify the Xen hypervisor to notify its
scheduling events to the guest OS through a kernel module. Our method does not need any kernel extension and all the measurements are performed in user mode.

Finally, there are monitoring services, such as CloudWatch~\cite{AWS2018CloudWatch} provided by AWS, that provides steal time monitoring under a given pricing model. The main constraint of tailored monitoring services, in addition to the economic cost, is it is not an agnostic solution, since it directly depends on the cloud provider itself. Therefore, developers need to adapt the software per every cloud provider under use to properly track the monitoring information.

\section{Main Idea}
\label{sec:algorithm}
\begin{figure}[th]
	\centering
	\includegraphics[width=0.45\textwidth]{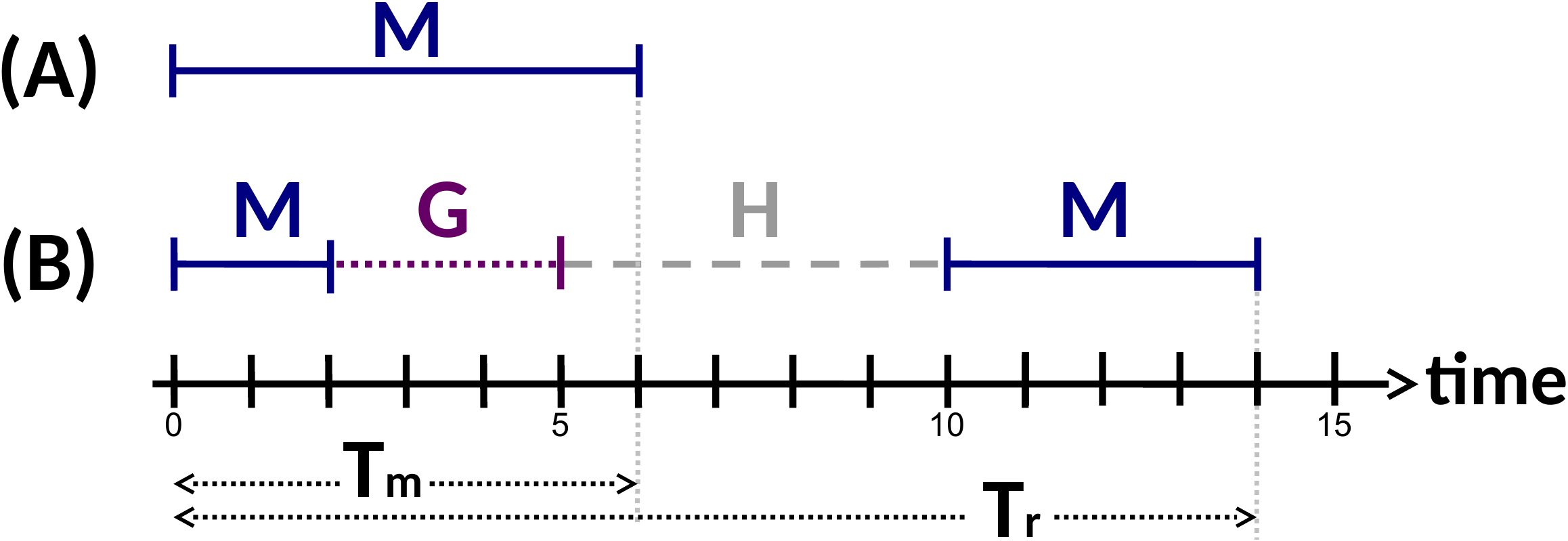}
	\caption{Execution times of microbenchmark (M) in (A) a non-virtualized isolated
	environment and in (B) a virtualized environment sharing resources with
	other applications in the same guest (G) and other VMs (H)}
	\label{fig:mainidea}
\end{figure}

Figure~\ref{fig:mainidea} shows the performance of a CPU bound application when it is executed in a non-virtualized isolated environment (A) and in a virtualized environment (B). In (A), the application (M) runs alone in a processor and it is not preempted by another process. So, $T_m$ is the minimal time the application needs to complete its execution. In (B) the application (M) runs in a virtualized environment along with other applications. In this case, the scheduler of the guest OS selects another application to execute (G), enlarging the time needed to execute (M). In addition, since the CPU time is multiplexed among VMs, the hypervisor can suspend the virtual machine to execute another VM (H), enlarging, even more, the time to completely execute the application. So, $T_r$ is the time to execute the application in (B) and is larger than (A) because it includes the interference of the other applications in the same VM and the the other co-scheduled VMs. The main idea in this paper is to estimate (H) having into account that (M) is known, since we use a synthetic microbenchmark with a calculable execution time, and (G) is also known since it corresponds to the CPU load of the guest OS.

Our approach consists in: (1) calculate the theoretical microbenchmark
(M) execution time $T_m$; and (2) periodically execute (M) inside
the VM to measure its real execution time $T_r$ and the load of the guest OS (G). As a result, we can estimate the time (H) that other VMs are executing in the physical processor.

As (M) must be obtainable either theoretically or
empirically, for this paper we implement a synthetic microbenchmark with a deterministic
behavior. Its code (Figure~\ref{fig:loopcode}) comprises a
sequence of dependent instructions, in this case \emph{adds}, using the same
register as source and destination to prevent any instruction reordering or
overlapping. The execution of this code presents a cycle per
instruction (CPI) of 1, and it is executed 10
million times to take representative measurements.

\begin{figure}
	\centering
\begin{verbatim}
__asm__ __volatile__ (
  "start:\n"
    "add %%rax, %%rax\n"
    ... repeat 50 times ...
    "add %%rax, %%rax\n"
    "sub $1,%%rcx\n"
    "jne start\n"
    :: "c"(spincount)
	);
\end{verbatim}
\caption{Loop microbenchmark source code}
\label{fig:loopcode}
\end{figure}

The theoretical execution time, $T_m$, of the microbenchmark is calculated as
the number of total
$instructions$ executed (in our case, 500 millions), multiplied by
the $CPI$ of the microbenchmark (1 cycle per instruction) and divided by the
frequency ($F$) of the processor (1.20GHz), as shown in
Formula~\eqref{eq:exectime}. Empirically it takes 420ms on average in the selected platform (see Section~\ref{sec:methodology}).

\begin{equation} \label{eq:exectime}
	T_m = \frac{instructions * CPI}{F}
\end{equation}

The delay in the microbenchmark execution time, $T_{(m+g)}$ in
Formula~\eqref{eq:execGMTime}, due to other applications running in the same VM,
such as (G) in Figure~\ref{fig:mainidea}, is calculated by combining the
theoretical time, $T_m$, with the $load$ of the
guest OS. This load is calculated as the number of runnable processes in the
guest multiplied by the averaged CPU load (0..1), ranging from an idle CPU to a totally
busy CPU executing applications. There are multiple tools that supply these
measurements.

\begin{equation}\label{eq:execGMTime}
	T_{(m+g)} = T_m * load
\end{equation}

Finally, the real execution time, $T_r$, of the microbenchmark is obtained empirically inside the VM.
The difference between $T_r$ and $T_{(m+g)}$ is the time other VMs are executing in the same host and thus, the \emph{steal time} in Formula~\eqref{eq:stealTime}.

\begin{equation}\label{eq:stealTime}
	steal time = T_r - T_{(m+g)}
\end{equation}


As it can be seen, this approach is platform agnostic, since its only requirements are: (1) the guest OS provides the measurement of real execution time and, (2) the load of the guest OS can be obtained, which are common features presented in all current operating systems.

\section{Experimental Framework}
\label{sec:methodology}
To evaluate our mechanism, we run a prototype of our proposal to measure the 
\emph{steal time} within the virtual machine in 4 different scenarios:
an idle VM, 25\%, 50\% and 100\% busy VM. This load is simulated through a
single process (see Section~\ref{sec:loadgenerator}). Besides, in
the same host machine we create a particular CPU
load pattern (see Section~\ref{sec:loadgenerator}) easily recognizable to
force resource contention.


\subsection{Platform}
The evaluation platform is a PC with an Intel Corei5-3320M @ 2.60GHz with
8GB RAM running a Linux Debian 8 operating system using a kernel 3.16. Although we do not explicitly use a hypervisor in this experimental environment, the scheduler of the host OS mimics the behavior of a real hypervisor in our simulated scenarios. On top of this, VirtualBox 4.3.36 is used to virtualize a single core 1GB of RAM machine running the same operating system. To obtain more stable results, the CPU frequency scaling and Turbo Boost features are disabled and the
frequency of the CPU is set to the lowest possible value (1.20GHz).

To avoid other processes interfering in the measurements, the VM and the load generator are pinned to the same core, and any remaining
processes in the host are excluded from this core using the Linux \emph{taskset}
command.


To obtain the execution time of the microbenchmark \emph{$T_r$} we use
the \emph{clock\_gettime} Linux system call using a monotonic clock.
Even if these measurements are Linux dependent, they can be easily ported to, \textit{e.g.} Microsoft Windows, by using the \emph{QueryPerformanceCounter} API.

\subsection{CPU load generator}
\label{sec:loadgenerator}
The CPU load generator is an application that generates an specific 
amount of CPU load average per second.
This generator is an iterative program with a CPU bound code followed by a
delay. The execution time of the code is known and 
the delay, defined by a factor, releases the CPU during a multiple of this execution time. 
The factor used in the delay  
determines the $load$ (Formula~\eqref{eq:execGMTime}) of the guest OS.


The CPU load generator is also used to simulate an easily recognizable pattern in the physical CPU to check if our
approach correctly measures the \emph{steal time}. This pattern consists in 
alternating idle (\textit{I}) and busy (\textit{B}) periods, 30 seconds each. The host presents 0\% CPU load during \textit{I}~periods. That is, there are no other VMs multiplexed. \textit{B}~periods simulate 100\% CPU load in the host, meaning the guest OS is multiplexed with another VM which demands total processor usage. Thus, \textit{B}~periods expose \textit{steal time}.

\section{Results}
\label{sec:results}
The figures in this section present the real execution time, $T_r$, solid lines, and \emph{steal time}, dashed lines, under different load scenarios. The
\emph{steal time} is obtained by measuring $T_r$ and applying Formula~\eqref{eq:stealTime}. In all figures, the y-axis
shows the time, in milliseconds, and the x-axis shows the different measurements performed. Along with the horizontal axis, we depict every idle and busy periods of the host.

\begin{figure}[th]
	\centering
	\includegraphics[width=0.42\textwidth]{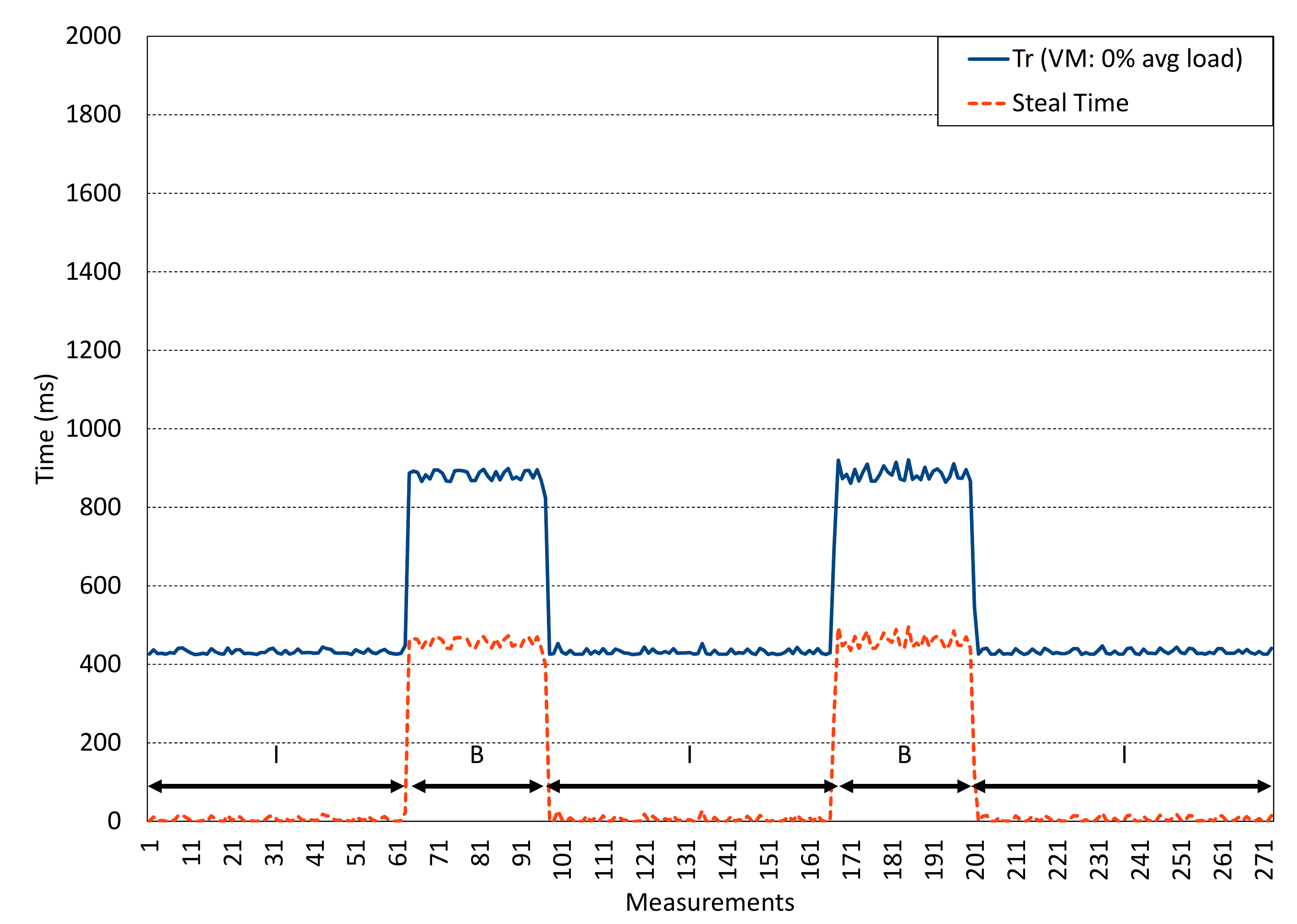}	
	\caption{Measurements inside an idle Virtual Machine}
	\label{fig:idlevm}
\end{figure}
Figure \ref{fig:idlevm} shows the execution time when there are no other processes
running in the guest OS. The microbenchmark takes around 430ms to execute when
the VM is alone. However, when other VMs demand 100\% CPU the host and the guest
OS are multiplexed, $B$~stages, and the application execution time is increased by 100\% accordingly. That is, busy periods present a steal time of 450ms on average. The execution time comes back to normal when the host goes idle,
$I$~periods. Although both busy and idle periods take 30 seconds each, it is interesting to note that
the number of measurements in $B$ stages are smaller than $I$ periods, because every $T_r$ takes longer.

\begin{figure}[th]
	\centering
	\includegraphics[width=0.42\textwidth]{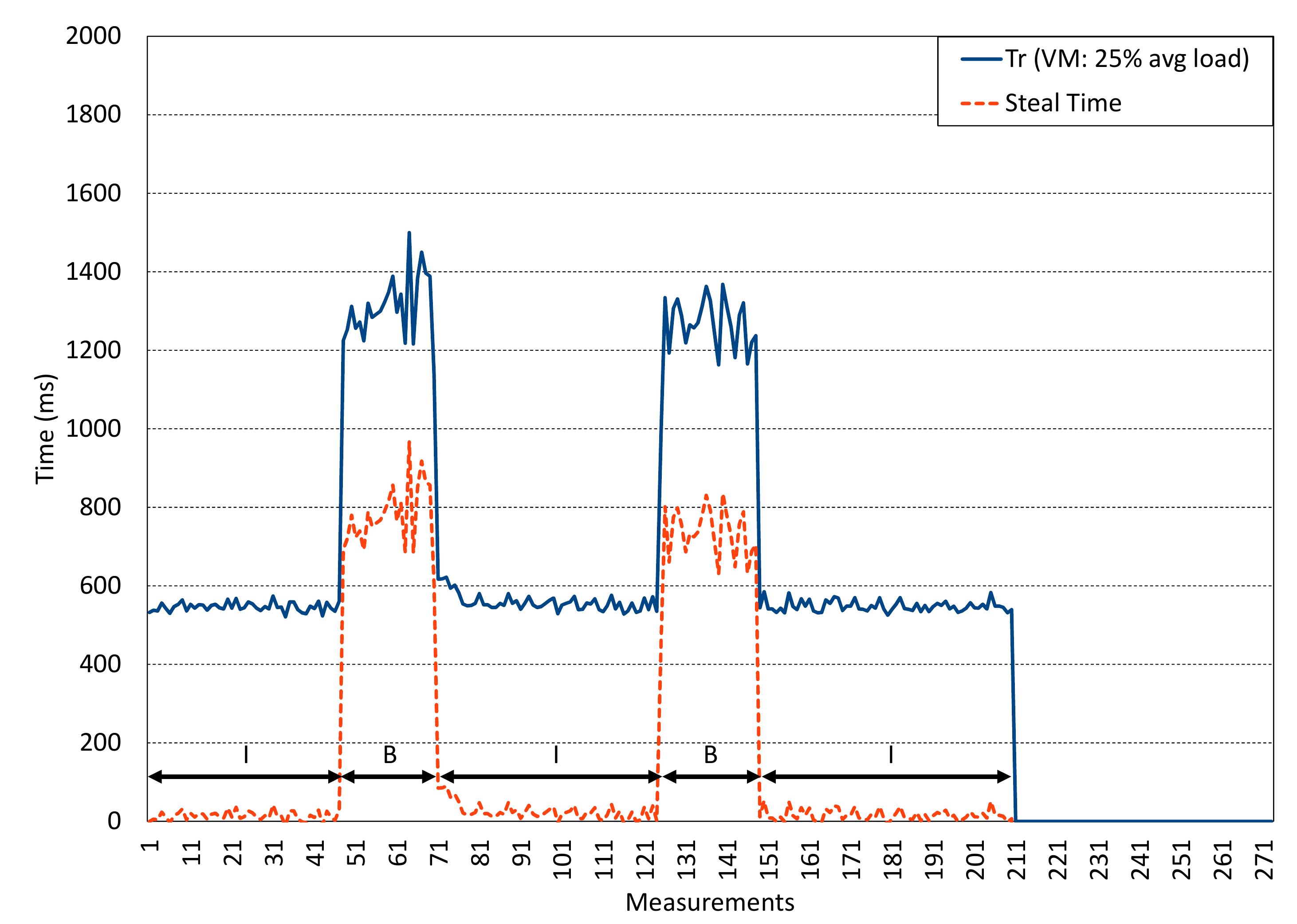}
	\caption{Measurements inside a Virtual Machine with 25\% load}
	\label{fig:busy25}
\end{figure}

When we add a 25\% CPU load inside the VM (Figure~\ref{fig:busy25}), we observe
spikes in the measurements with higher variations in busy periods. $T_r$
increases up to 545ms on average during $I$~periods, about 26.7\% compared to
an idle VM execution times. The impact is higher during $B$ periods, about 49\%
compared to idle VM. In fact, steal time also increases up to 735ms on average.
Besides, as the load in the guest ($T_{m+g}$) increases the real execution time, the number of measurements in both $I$ and $B$ periods are accordingly smaller compared to Figure~\ref{fig:idlevm}. 

\begin{figure}[th]
	\centering
	\includegraphics[width=0.42\textwidth]{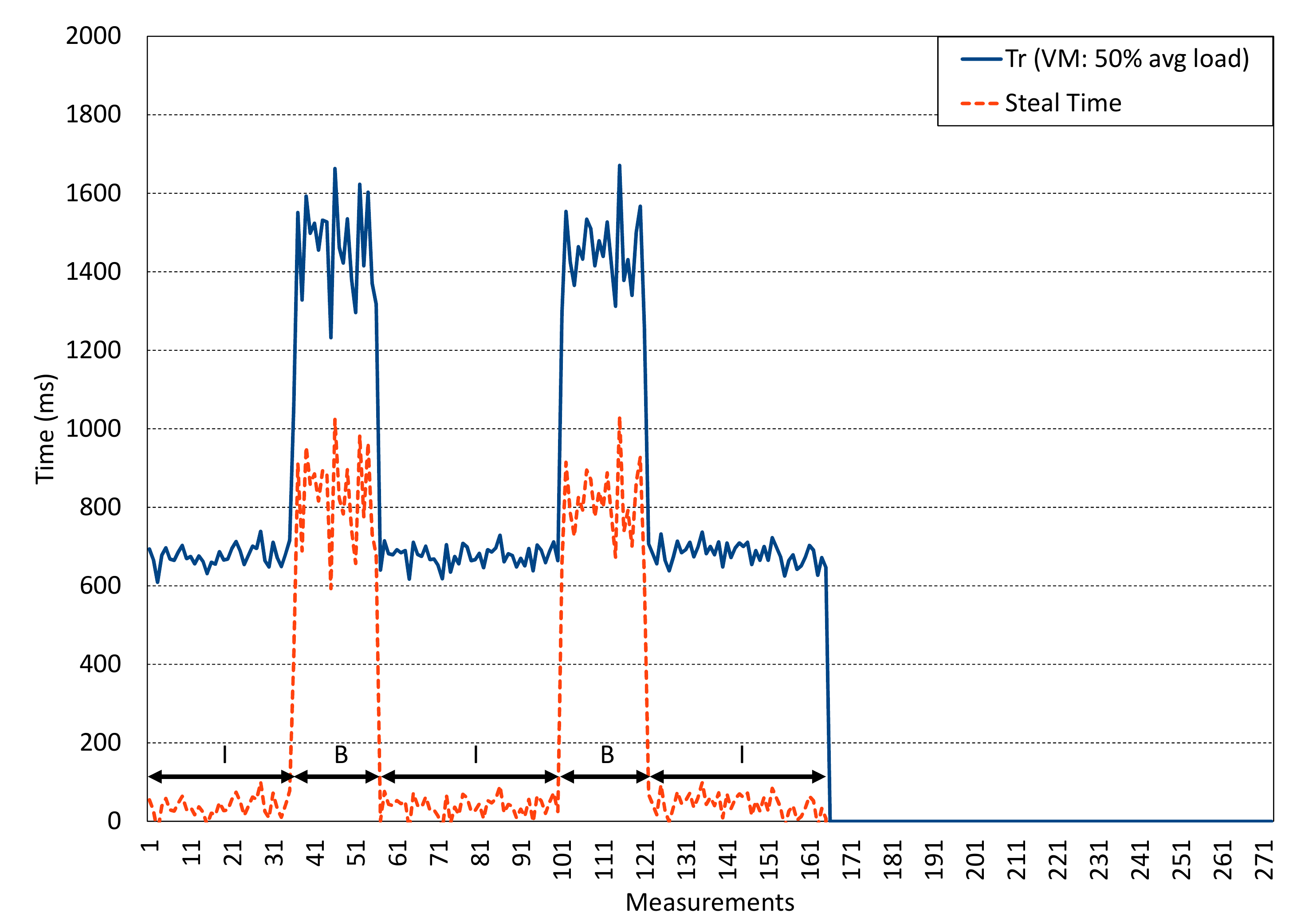}	
	\caption{Measurements inside a Virtual Machine with 50\% load}
	\label{fig:busy50}
\end{figure}
\begin{figure}[th]
	\centering
	\includegraphics[width=0.42\textwidth]{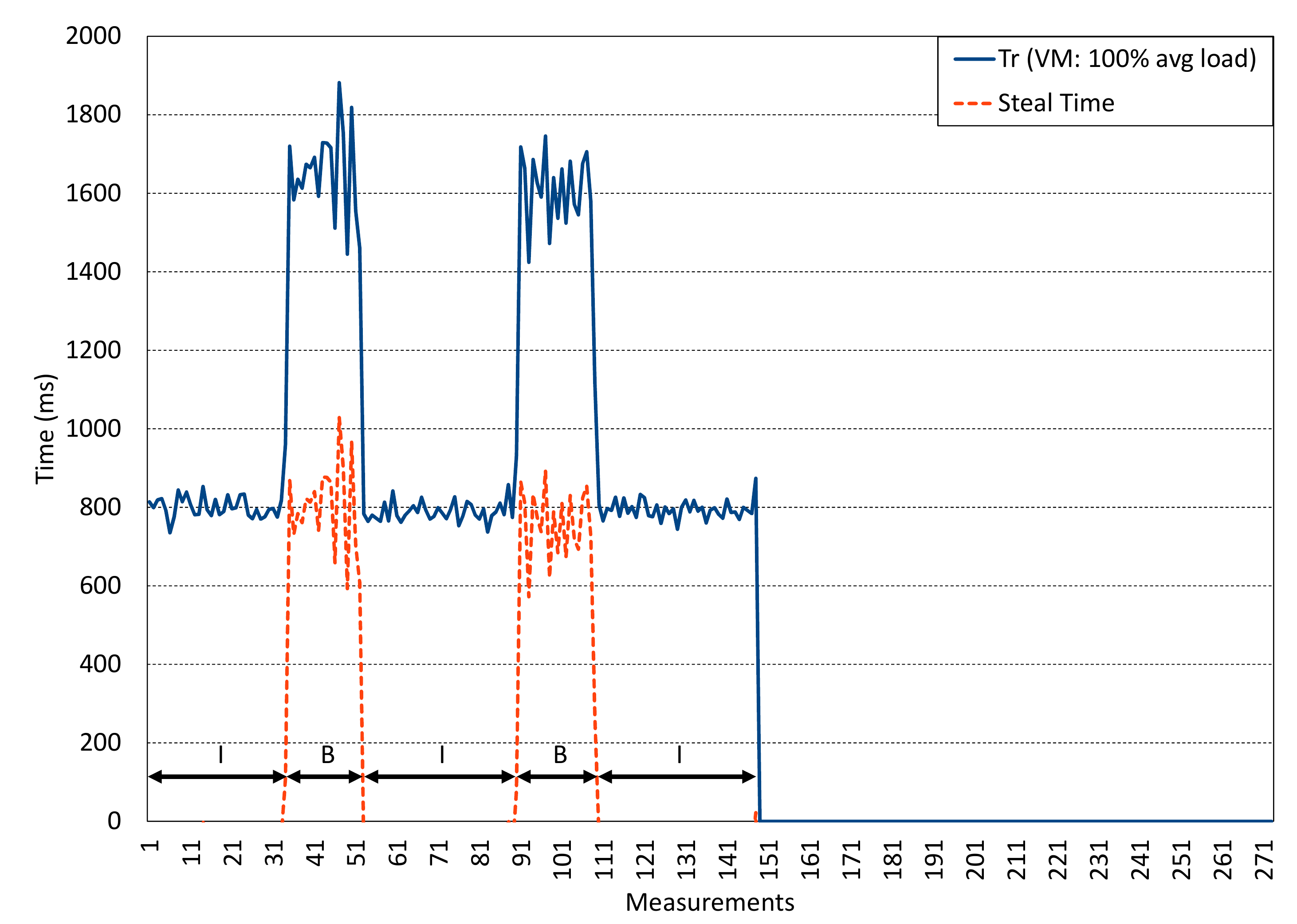}	
	\caption{Measurements inside a Virtual Machine with 100\% load}
	\label{fig:busy100}
\end{figure}
Figures \ref{fig:busy50} and \ref{fig:busy100}
show more noticeable time spikes with a VM load of
50\% and 100\%, respectively. Under these scenarios, the $T_r$
presents a direct impact on execution time increments due to the larger load in the guest OS and accordingly to $T_{(m+g)}$ increments.
On the other side, the steal time slightly
increases up to 775ms and 842ms on average, respectively.

\section{Future Work}
\label{sec:futurework}
Even though the results show that our approach differentiates the execution time, the CPU load from inside a VM, and the steal time, our
current work has two major limitations. The first one is that only one single
core VMs are considered. The second one is that this approach considers a fixed CPU clock frequency.

As future work we plan to take into account multi--core architectures and the effects of variations on clock frequency due to CPU throttling, as its value is key to calculate this steal time. Also, we plan to test our approach in a more realistic scenario with a real hypervisor in public elastic cloud infrastructures including dynamic load in guest OSes.

\section{Conclusion}
\label{sec:conclusions}

The real execution time of a program measured in a VM takes into
account the time that the hypervisor has suspended the virtual machine, the
steal time. This steal time is visible from the guest OS if it is tightly coupled with the virtualization layer. 
In other scenarios, like hardware-assisted virtual machines and not coupled guest OS-hypervisor setups, this measurement is not available what leads to an incorrect evaluation of the performance of applications. 

This work describes a platform agnostic approach to measure the steal time in a guest OS without any
cooperation from the virtualization layer.
Preliminary results show the feasibility of this approach and that steal
time is effectively computed in a guest Linux OS running on a hardware-assisted
virtual machine.

\section*{Acknowledgments}
This work has been partially supported by the AWS Cloud Credits for Research program. The authors greatly appreciate the recognition from the Generalitat de Catalunya of VIRTUOS as Emergent Research Group (2017--SGR--0962).



\end{document}